\documentclass[structabstract]{aa}  
\usepackage{graphicx}
\usepackage{lscape}
\usepackage{soul}
\usepackage{amssymb}
\usepackage{txfonts}
\begin{document}
   \title{The submillimeter spectrum of deuterated glycolaldehydes}

   %\subtitle{}

   \author{A. Bouchez
          \inst{1,2,3}
          \and
            L. Margul\`{e}s\inst{3}
            \and
            R.~A. Motiyenko\inst{3}
            \and
            J-C. Guillemin\inst{4}
            \and
            A. Walters\inst{1,2}
            \and
            S. Bottinelli\inst{1,2}
            \and
            C. Ceccarelli\inst{5}
            \and
            C. Kahane\inst{5}}
   \institute{Universit\'e de Toulouse; UPS-OMP; IRAP; Toulouse, France.\\
              \email{abouchez@irap.omp.eu}  
         \and
			 CNRS; IRAP; 9 Av. colonel Roche, BP 44346, F-31028 Toulouse cedex 4, France.
         \and             
             Laboratoire de Physique des Lasers, Atomes, et Mol\'ecules, UMR CNRS 8523, 
               Universit\'e de Lille I, F-59655 Villeneuve d'Ascq C\'edex, France
		 \and           
             Sciences Chimiques de Rennes, Ecole Nationale Sup\'erieure de Chimie de Rennes, CNRS, UMR 6226, Avenue du
         G\'en\'eral Leclerc, CS 50837, 35708 Rennes Cedex 7, France
         \and
             IPAG, Universit\'e Joseph Fourier, CNRS, BP 53 F-38041, Grenoble cedex 9, France}
   \date{}
% \abstract{}{}{}{}{} 
% 5 {} token are mandatory
 
 \abstract{Glycolaldehyde, a sugar-related interstellar prebiotic molecule, has recently been detected in two star-forming regions, Sgr B2(N) and G31.41+0.31. The detection of this new species increased the list of complex organic molecules detected in the interstellar medium (ISM) and adds another level to the chemical complexity present in space. Besides, this kind of organic molecule is important because it is directly linked to the origin of life. For many years, astronomers have been struggling to understand the origin of this high chemical complexity in the ISM. The study of deuteration may provide crucial hints.}
   {In this context, we have measured the spectra of deuterated isotopologues of glycolaldehyde in the laboratory: the three mono-deuterated ones (CH$_2$OD-CHO, CHDOH-CHO and CH$_2$OH-CDO) and one dideuterated derivative (CHDOH-CDO) in the ground vibrational state.}
   {Previous laboratory work on the D-isotopologues of glycolaldehyde was restricted to less than 26 GHz. We used a solid-state submillimeter-wave spectrometer in Lille with an accuracy for isolated lines better than 30 kHz to acquire new spectroscopic data between 150 and 630 GHz and employed the ASFIT and SPCAT programs for analysis.}
   {We measured around 900 new lines for each isotopologue and determined spectroscopic parameters. This allows an accurate prediction in the ALMA range up to 850 GHz}
   {This treatment meets the needs for a first astrophysical research, for which we provide an appropriate set of predictions.}


   \maketitle
%
%________________________________________________________________

\section{Introduction}

   A search for various major isotopologues of key species in the interstellar medium (ISM) is still hampered by a scarcity of reliable spectroscopic data from laboratory measurements, which has also been the case for deuterated glycolaldehyde. However, studying the spectra and abundances of isotopologues in the ISM can be a valuable tool in chemical and physical modelling. In particular, a study of the isotopologues of known organic molecules can provide information for understanding molecular complexity in star-forming regions (SFRs). The detection of deuterated molecules has a strong impact for understanding and modelling the formation and deuteration process of molecules. This is especially interesting for complex molecules for which it may allow differentiating the gas-phase and solid-state formation pathways.

	Glycolaldehyde is the smallest possible molecule that contains both an aldehyde group and a hydroxyl group and this dimer of formaldehyde can be considered as the simplest sugar. It is an  isomer of methyl formate (HCOOCH$_3$) and acetic acid (CH$_3$COOH), and the relative abundance of these isomers are two of the three exceptions of the minimum energy principle (\cite{2009ApJ...696L.133L}); the most stable isomer, acetic acid, should be the most abundant, but in all astronomical sources where it was searched for, methyl formate is much more abundant, and acetic acid and glycolaldehyde have about the same abundance. These exceptions could be due to the kinetic or abundance effect, or to the existence of different routes of formation with no intermediate in common or to specific depletion on the grains of one isomer with respect to the others.
	
	Glycolaldehyde is a probable prebiotic molecule and a key intermediate in the formose reaction, i.e the formation from formaldehyde of sugars containing three, four, or five C-atoms.
It has been detected in the star-forming region Sgr B2(N) by \cite{2000ApJ...540L.107H} from 2 and 3 mm single-dish surveys, with an estimated rotational temperature (T$_{\rm rot}$) of 35K from 34 lines detected. \cite{2009ApJ...690L..93B} reported identification of three lines towards the hot core G31.41 +0.31 using the Plateau de Bure Interferometer. Their analysis indicated that emission comes from the hottest ($\geq$ 300K) and the densest ($\geq$ 2 $\times$ 10$^8$ cm$^{-3}$) region closest to the protostar. Although some transitions of the parent molecule have been observed, no investigation of the isotopologues has yet been reported because of (i) the lack of spectroscopic data and (ii) the expected weakness of their emission. However, observations like this would help us to understand the formation pathway of the parent molecules, which is currently unclear.
Indeed, several gas-phase and solid-phase routes of formation of glycolaldehyde have been proposed. The possible gas-phase pathway is the formose reaction (\cite{2007AsBio...7..433J}): protonated formaldehyde CH$_2$OH$^+$ can react with formaldehyde to yield glycoladehyde. However, a recent study by \cite{Simakov} compromises this formation route: the authors reported that protonated formaldehyde does not lead to protonated glycolaldehyde. Other recent studies (e.g. \cite{2010A&A...514A..83H}) have shown that dissociative recombination reactions of a protonated molecule HX$^+$ do not lead to the stable molecule X for the most part.
Hence a more probable route of formation is via the reaction of methanol and carbon monoxide on interstellar ices (\cite{2007ApJ...661..899B}):

\begin{center}
CH$_3$OH + CO $\rightarrow$ CH$_2$OH + HCO $\rightarrow$ OHCH$_2$-CHO.
\end{center}

To date, no identification of the deutero-isotopologues has been reported, but a comparison of D/H ratios may be valuable for understanding the formation of glycolaldehyde. Indeed, in hot cores and hot corinos the deuteration fraction of complex organic molecules is much greater than the cosmic D/H ratio (10$^{-5}$; \cite{Oliveira}) because it is around 10$^{-3}$. 
This ratio is more studied in hot corinos where the process of deuteration is considered to be overabundant (detected molecules can be doubly or triply deuterated, as described for example by \cite{Ceccarelli1998}). This is used to determine the formation pathway as shown by \cite{2011ApJ...741L..34C} and \cite{coutens2011}, with the formation of water and formaldehyde. However, the process could be similar for hot cores, as shown by \cite{1990ApJ...362L..29T} or recently by \cite{2011A&A...528L..13R}. Therefore we recently made extensive laboratory measurements to be able to search for the millimeter-wave spectrum of D-isotopologues of glycolaldehyde in the ISM. These measurements are particularly relevant to ALMA because of its high sensitivity and high spatial resolution (that could, for example, reduce the spectral line confusion). However, it might also be possible to identify these isotopologues using sensitive single-dish instruments such as the IRAM-30m.

\begin{figure}[h!]
\includegraphics[width=\columnwidth]{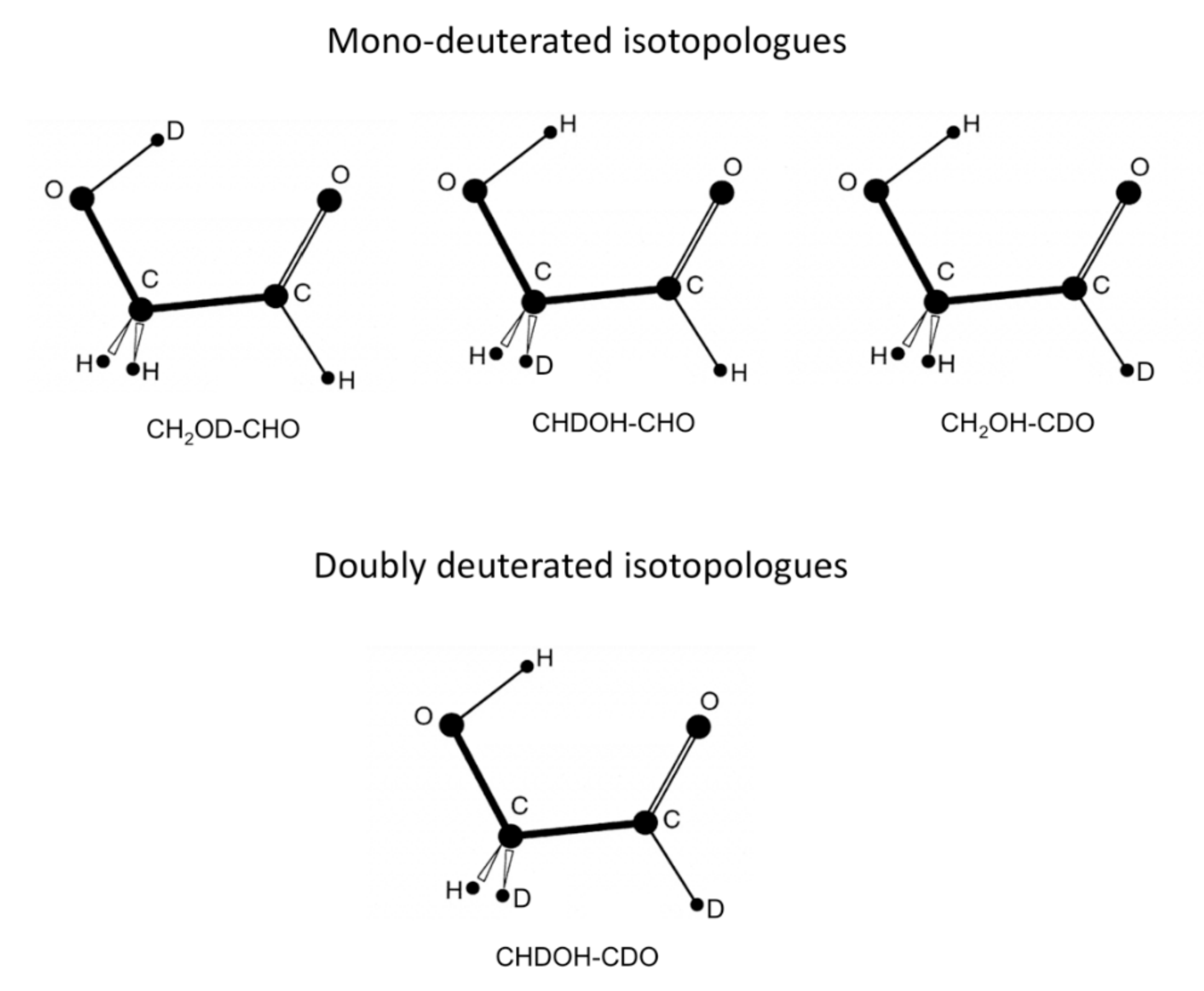}
\caption{Representation of all mono-deuterated isotopologues of glycolaldehyde and the doubly deuterated one.}
\label{mol}
\end{figure}

The first microwave spectrum measurements of the main isotopologue of glycolaldehyde were reported by \cite{1970JMoSt...5..205M}, who obtained the pure rotational and centrifugal distortion constants of the ground state and three vibrational excited states, as well. In addition, in 1973 they were able to determine the electric dipole moment as $\mu_b$ = 2.33~D and $\mu_a$ = 0.26~D. Then \cite{2001ApJS..134..319B} extended millimeter and submillimeter assignments for the ground vibrational state from 128 to 354 GHz. A subsequent study by \cite{2005ApJS..158..188W} included assignments for the ground state and three excited vibrational states up to 354 GHz as well as a partition function analysis. Finally, \cite{2010ApJ...723..845C} reported the spectrum of the ground vibrational state in selected regions of the submillimeter region up to 1.2 THz.
Less information is available on the other isotopologues of glycolaldehyde. The only published measurements for the deuterium isotopologues are from \cite{1971JMoSt...7..101M}, whose measurements were restricted to between 12 and 26 GHz. Considering the importance of understanding the formation of this molecule and the development of (sub)millimeter observatories, new data are needed, which is why we measured the spectra of all mono-deuterated isotopologues (CH$_2$OD-CHO, CHDOH-CHO, CH$_2$OH-CDO, cf. figure~\ref{mol}) of glycolaldehyde and one doubly deuterated one (CHDOH-CDO, cf. figure~\ref{mol}) between 150 and 630 GHz in the laboratory.

%__________________________________________________________________

\section{Experiments}

The submillimeter-wave measurements (150-630 GHz) were performed using the Lille spectrometer (\cite{Motiyenko2010}) based only on solid-state sources. The frequency of an Agilent synthesizer E8257D (12.5-17.5 GHz) was first multiplied by six and amplified by a Spacek active sextupler, providing an output power of +15dBm in the W-band range (75-110 GHz). This power is high enough to use passive Schottky multipliers (x2, x3, x5, x3x2) from Virginia Diodes Inc in the next stage of the frequency multiplication chain. As a detector we used an InSb liquid He-cooled bolometer from QMC Instruments Ltd. To improve the sensitivity of the spectrometer, the sources were frequency modulated at 10 kHz and phase-sensitive detection at 2f was employed. The absorption cell is a stainless-steel tube (6 cm diameter, 220 cm long). The sample pressure during measurements was about 1.5 Pa (15 $\mu$bar) and the linewidth was limited by Doppler broadening. Measurements were performed at room temperature. The accuracy of the measurements for isolated lines is estimated to be better than 30 kHz. However, if the lines were blended or had a poor signal-to-noise ratio, they were given a weight of 100 or even 200 kHz.\\

		The synthesis procedure was as follows:

Synthesis of deuterated glycolaldehydes:
1,4-Dioxane-2,5-diol and dihydroxyfumaric acid hydrate were purchased from Aldrich and used without additional purification.

Synthesis of DOCH$_2$CHO:
1,4-Dioxane-2,5-diol (5 g) was dissolved in deuterated water (10 mL) and heated to 70$\degr$C for 1h. The mixture was lyophylisated to obtain a solid sample of DOCH$_2$CHO  with about 25$\%$ isotopic purity.

Synthesis of HOCHD-CHO, HOCH$_2$-CDO and HOCHD-CDO:
The synthesis reported by \cite{1973JMoSt..16..259M} was used starting from dihydroxyfumaric acid hydrate dissolved in about 80$\%$ D$_2$O. Decarboxylation in dry pyridine as described by Powers et al. gave an about 2:1:1 mixture of HOCHD-CHO, HOCH$_2$-CDO and HOCHD-CDO. Small amounts of residual pyridine, a compound more volatile than glycolaldehyde, led to relatively intense signals in the microwave spectrum.

%___________________________________________________________________

\section{Analysis}

Glycolaldehyde is a prolate asymmetric top molecule ($\kappa=-0.7$). Therefore we used the Watson A-reduction Hamiltonian in the $I^r$ representation in the analysis of its spectra. Spectral data were fitted using the ASFIT programme\footnote{Kisiel, PROSPE, http://www.ifpan.edu.pl/~kisiel/prospe.htm}. Predictions were made using the SPCAT programme (\cite{1998JQSRT..60..883P}) and the parameter set resulting from ASFIT and the dipole moments determined by \cite{1973JMoSt..16..259M}. \\

\begin{figure}[h!]
\includegraphics[width=\columnwidth]{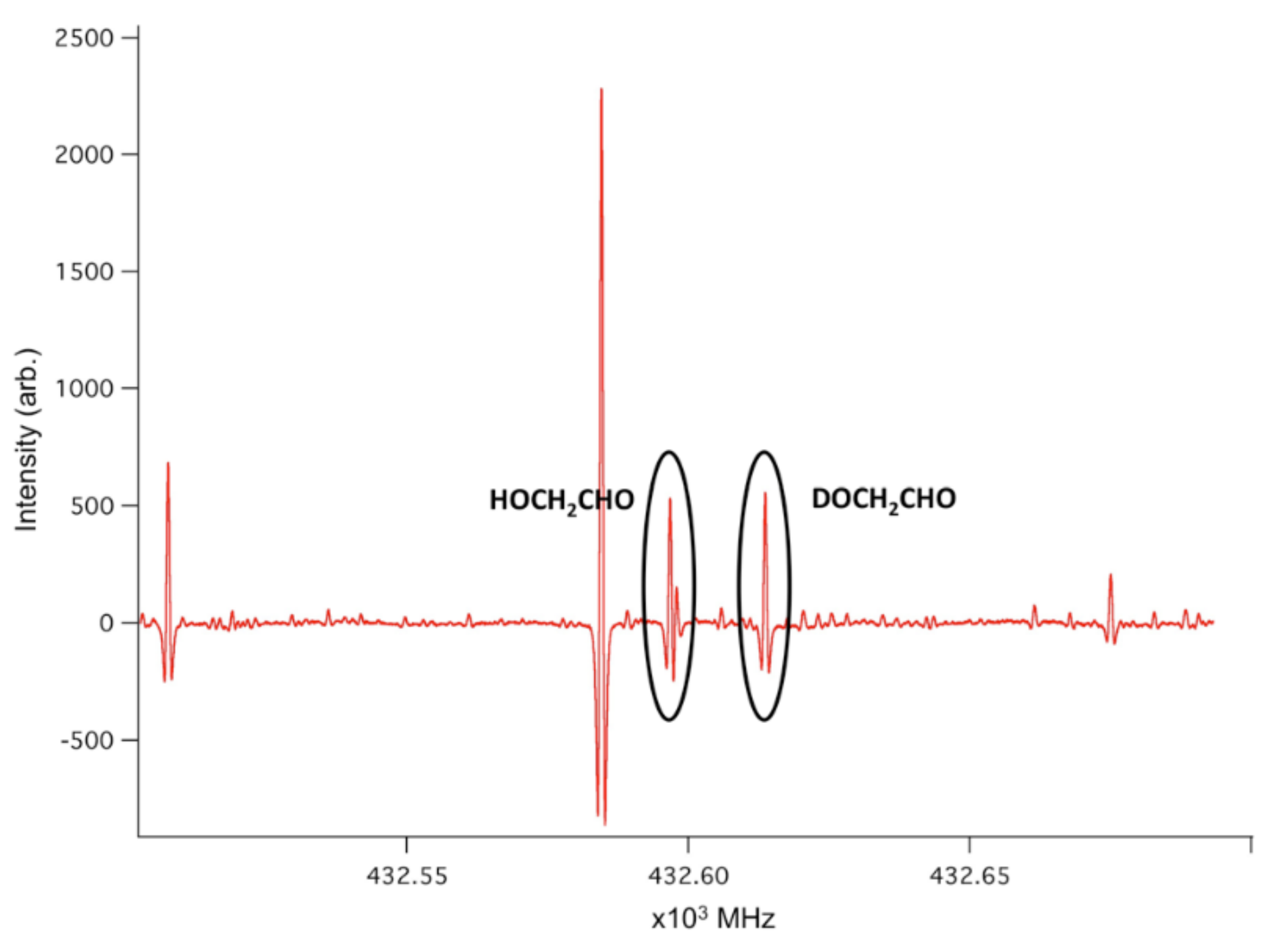}
\caption{Part of the spectrum of the first sample containing HOCH$_2$CHO (transition 39$_{7,32}$-38$_{8,31}$ and intensity $\alpha$=1.37*10$^{-4}$ cm$^{-1}$) in the first excited vibrational state and DOCH$_2$CHO (transition 38$_{7,32}$-37$_{6,31}$ and intensity $\alpha$=5.83*10$^{-4}$ cm$^{-1}$).}
\label{fig1}
\end{figure}

The first species measured and analysed was DOCH$_2$CHO with the enriched sample containing only this isotopologue and the parent species with an estimated proportion of 25$\%$ of OD as shown in figure~\ref{fig1}. 

Initial predictions were made using lines reported in \cite{1971JMoSt...7..101M} between 12 and 26 GHz. First, we identified the strong R-branch transitions with K$_a$ = 0 and 1 between 150 and 310 GHz up to J=31. The low J lines were easily recognised doublets (K$_a$ selection rules: 1-0 \& 0-1) with systematic decrease in the split towards higher J and visible addition of the intensities as the components merge. All transitions were shifted systematically from the predictions up to a maximum of around 10 MHz at J = 20. The newly identified lines were then included in the fit. Next we identified lines by systematically increasing K$_a$ up to K'$_a$ =11 and finally measured all other identifiable R-branch lines (up to J’=59 and K'$_a$= 18). We then identified three Q branches and assigned all lines possible. The same procedure was then repeated for a second frequency band between 400 and 630 GHz. Finally, we assigned some P-branch transitions. Since these lines were much weaker, assignment was only possible in the region around 300 GHz where the spectrometer is most sensitive. Because the strong \textit{b}-component of the dipole moment (2.33 D) is much stronger than the \textit{a}-component (0.26 D), all newly assigned lines are \textit{b}-type transitions.

To check the relative weight of the Lille and centimeterwave data sets, the method of iteratively reweighted least-squares (IRLSQ) of \cite{Margules2010} was used. As a result, 63 of the 65 lines from \cite{1971JMoSt...7..101M} were rejected from the fit. We found a systematic deviation of 150 kHz for these measurements, and consequently they were excluded from the final fit. The same kind of deviation was also found for the data on other deuterated species available from \cite{1971JMoSt...7..101M}. The final fit is given in the supplementary Table~\ref{table2}. It contains 889 new identified lines (1433 transitions, because there are 355 unresolved doublets and 63 quadruplets) up to J’ = 64 and K'$_a$ = 37.

We were able to determine all sextic centrifugal distortion parameters. Including some octic parameters made no significant improvement and consequently they were excluded from the final fit. The rms deviation of the fit is 27 kHz  with a weighted rms of 0.90, which indicates that the experimental uncertainty was correctly estimated. Table~\ref{table1} gives the rotational and centrifugal distortion
parameters. 

The second studied sample contained the three isotopologues CHDOH-CHO, CH$_2$OH-CDO, and CHDOH-CDO with an abundance ratio of ${2:1:1}$. The analysis was carried out for each isotopologue independently, as previously described for CH$_2$ODCHO, and all newly assigned lines are \textit{b}-type transitions, too. However, the analysis was complicated by the presence of three different isotopologues and the presence of many pyridine lines resulting from the synthesis procedure.

\begin{figure}[h!]
\includegraphics[width=\columnwidth]{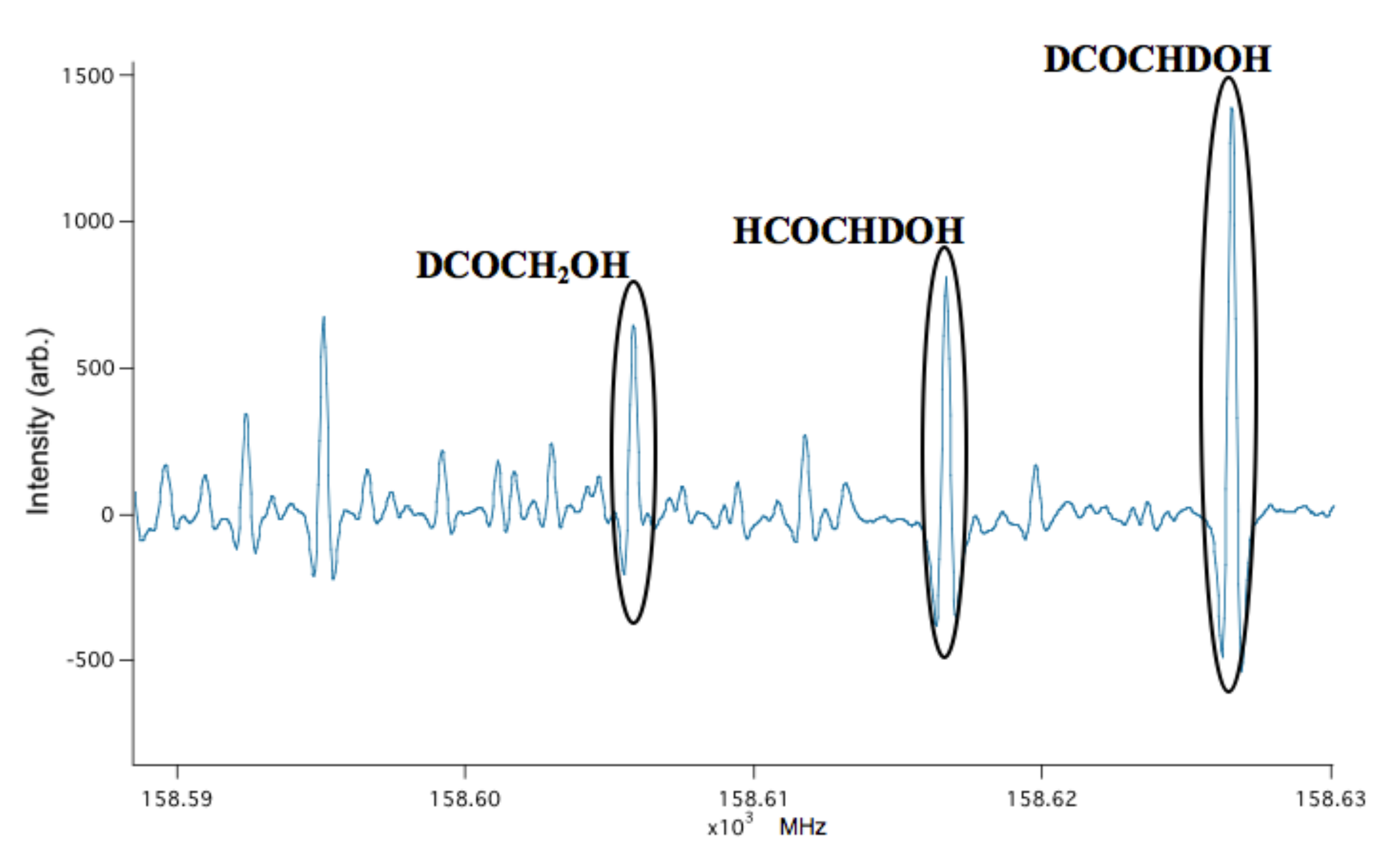}
\caption{Part of the spectrum of the second sample containing CHDOH-CHO (transition 23$_{3,20}$-23$_{2,21}$ and intensity $\alpha$=1.62*10$^{-4}$ cm$^{-1}$), CH$_2$OH-CDO (transition 7$_{4,3}$-6$_{3,4}$ and intensity $\alpha$=1.90*10$^{-4}$ cm$^{-1}$) and CHDOH-CDO (transition 33$_{10,23}$-33$_{9,22}$ and intensity $\alpha$=2.01*10$^{-4}$ cm$^{-1}$).}
\label{fig2}
\end{figure}
   
For CHDOH-CHO a total of 945 new lines were identified (1340 transitions because there are 179 unresolved doublets and 72 quadruplets) up to J’ = 64 and K'$_a$ = 27.\\
	For CHDOH-CDO we obtained 916 new identified lines (1368 transitions besauce there are 197 unresolved doublets and 85 quadruplets) up to J’ = 67 and K'$_a$ = 29.\\
	Finally for CH$_2$OH-CDO we obtained 858 new identified lines (1268 transitions because there are 203 unresolved doublets and 69 quadruplets) up to J’ = 65 and K'$_a$ = 24.

Table~\ref{table1} gives the rotational and centrifugal distortion parameters determined for these three other isotopologues. We used the same parameter set for each isotopologue that was determined by trial as best (i.e, lowest rms with minimal number of parameters). All quartic and sextic centrifugal terms were determined. The uncertainties on the parameters and the weighted rms of the fits (0.87, 0.87 and 0.90 for CHDOH-CHO, CHDOH-CDO, CH$_2$OH-CDO respectively) are similar for all isotopologues. \\

\begin{table*}[t!]
\begin{center}
\caption{Determined rotational and centrifugal distortion parameters.}
\label{table1}
\begin{tabular}{|l|r|r|r|r|} 
\hline\hline             
Parameters & CH$_2$OD-CHO  & CHDOH-CHO & CHDOH-CDO & CH$_2$OH-CDO\\
\hline
        &                                  &                     &                          &                     \\
A /MHz &                17490.68230 (18)   &16987.80362    (18)  & 15862.45361 (17)         &   17150.99706 (21)  \\  
B /MHz &                6499.72649 (12)   & 6385.430683    (91)  &  6233.230843 (90)        &    6362.87482 (10)  \\  
C /MHz &               4882.948293 (76)  & 4843.811766    (70)   &  4663.584048 (65)        &    4778.922377 (73) \\  
$\Delta_J$ /kHz &       6.349256 (81)  &    5.622519    (55)     &     4.988629 (52)        &       5.471924 (59) \\  
$\Delta_{JK}$ /kHz &  -17.93851 (31)   &  -15.51877    (26)      &   -13.56931 (22)         &     -16.70029 (31)  \\  
$\Delta_K$ /kHz &      37.67004 (34)   &   35.90235    (46)      &    28.82809 (37)         &      36.52446 (81)  \\  
$\delta_J$ /kHz &       1.898798 (40)  &    1.656979    (23)     &     1.508496 (23)        &       1.650572 (26) \\  
$\delta_K$ /kHz &       8.81685 (86)   &    7.59317    (44)      &     6.10186 (36)         &       7.30024 (46)  \\  
$\Phi_J$ /mHz &        -0.008080 (30)  &   -0.006612    (17)     &    -0.005815 (16)        &      -0.006307 (18) \\  
$\Phi_{JK}$ /Hz &       0.15452 (42)   &    0.13274    (27)      &     0.10725 (21)         &       0.12995 (28)  \\  
$\Phi_{KJ}$ /Hz &      -0.6494 (15)    &   -0.5728    (10)       &    -0.43733 (80)         &      -0.5700 (10)   \\  
$\Phi_K$ /Hz &          0.7573 (12)    &    0.71373    (86)      &     0.50818 (67)         &       0.7048 (12)   \\  
$\phi_J$ /mHz &        -0.002662 (15)  &   -0.0022434    (84)    &    -0.0020597 (82)       &      -0.0021147 (88)\\  
$\phi_{JK}$ /mHz &     -0.03007 (40)   &   -0.02255    (23)      &    -0.01534 (19)         &      -0.01999 (22)  \\  
$\phi_K$ /Hz &          0.2636 (47)    &    0.1945    (29)       &     0.1254 (21)          &       0.1777 (26)   \\  
 &&&&\\                                                                                                          
 J$_{MAX}$, K$_{a MAX}$ &         64, 37    &               64, 27  &               67, 29  &             65, 24\\  
 Freq$_{MAX}$ /MHz      &     626777.73     &            629356.35  &           626675.43   &         623233.58 \\  
 nbr of transitions     &            889    &                   945 &                916    &               858 \\  
 $\sigma_{FIT}$ /kHz    &             27    &                    26 &                  26   &                 27\\  
\hline                                                                                                                                                                                          
\end{tabular}
\end{center}
\end{table*}

A full list of measurements is given in Tables~\ref{table2}, 3, 4 and 5 for CH$_2$OD-CHO, CHDOH-CHO, CH$_2$OH-CDO, and CHDOH-CDO, respectively. These tables give the rotational quantum numbers, the observed frequencies, the residuals (obs-cal), and the assumed experimental uncertainties. Blended transitions are treated by fitting the intensity-averaged frequency, and this weighting is also given in the tables. Lines with obvious experimental problems are not included. (Only the first 12 lines of Table~\ref{table2} appear in the paper edition; for a complete version of all species, see the electronic edition or http://cdsarc.u-strasbg.fr/cgi-bin/VizieR?-source=J/A+A/Vol/Num)

\begin{table*}[h!]
\centering
\label{table2}
\caption{Assigned transitions for the ground state of CH$_2$OD-CHO.}\label{table2}
\begin{tabular}{cccccccccc} 
\hline\hline             
J' & K'$_a$ & K'$_c$ & J" & K"$_a$ & K"$_c$ & Frequency & Obs. - Calc. & Uncertainty & Intensity weighting \\
  &  &  &  &  &  & (MHz) & (MHz) & (MHz) & for blended lines\\
  \hline
  25 & 11 & 14  &  25 & 10 & 15   &  242089.1640   &  -0.0346 &  0.0300   &        ...       \\
  25 & 11 & 15  &  25 & 10 & 16   &  242098.7040   &  -0.0043 &  0.0300   &        ...       \\
  24 & 11 & 13  &  24 & 10 & 14   &  242884.5120   &   0.0257 &  0.0300   &        ...       \\
  24 & 11 & 14  &  24 & 10 & 15   &  242888.6880   &  -0.0039 &  0.0300   &        ...       \\
  23 & 11 & 12  &  23 & 10 & 13   &  243577.9080   &   0.0077 &  0.0300   &        ...       \\
  23 & 11 & 13  &  23 & 10 & 14   &  243579.6720   &  -0.0094 &  0.0300   &        ...       \\
  22 & 11 & 11  &  22 & 10 & 12   &  244180.5840   &  -0.0272 &  0.0300   &        ...       \\
  22 & 11 & 12  &  22 & 10 & 13   &  244181.3400   &   0.0094 &  0.0300   &        ...       \\
  21 & 11 & 10  &  21 & 10 & 11   &  244702.4760   &   0.0192 &  0.0300   &        5.0E-01\\
  21 & 11 & 11  &  21 & 10 & 12   &  244702.4760   &   0.0192 &  0.0300   &        5.0E-01\\
  20 & 11 &  9  &  20 & 10 & 10   &  245151.7200   &   0.0551 &  0.0300   &        5.0E-01\\
  20 & 11 & 10  &  20 & 10 & 11   &  245151.7200   &   0.0551 &  0.0300   &        5.0E-01 \\
   \hline
\end{tabular}
\tablefoot{This table and those of other isotopologues are available in their entirety in the electronic edition in the online journal: http://cdsarc.u-strasbg.fr/cgi-bin/VizieR?-source=J/A+A/Vol/Num. A portion is shown here for guidance regarding its form and content.}
\end{table*}

%_____________________________________________________________________
\section{Conclusion}

The rotational spectra of three mono-deuterated glycolaldehyde isotopologues and a doubly deuterated one in the ground vibrational state have been characterised up to 630 GHz. This frequency range is appropriate for a first astronomical identification. Most importantly for an astrophysical research, the new measurements provide complete and precise predictions in the range of one to several hundred GHz, which is the optimum range for detection, taking into account the supposed temperature ($\geq$ 35K see introduction) and instrumental sensitivity. It is essential to be able to search for transitions over a wide frequency range not only to confirm the detection, but also because of spectral cluttering from other species. The measurements at higher frequency ensure the reliabilty of the predictions, but may also be useful in their own right, especially if glycolaldeyde should be detected at higher temperatures.
The increased sensitivity provided by a new generation of radio-astronomical instruments (e.g., upgrading at the IRAM-30m) and by new facilities such as the ALMA interferometer renders the prospect of detecting new deuterated isotopologues in astrophysical environments a realistic one. Using the recently measured data, we have therefore started to search for deuterated glycolaldehyde in existing spectral surveys of hot-core sources and plan new observations.

\begin{acknowledgements}
  This work is supported by the contract ANR-08-BLAN-0225 and by the Programme National de Physico-Chimie du Milieu Interstellaire (PCMI-CNRS) and J.-C. G. thanks the Centre National d'Etudes Spatiales (CNES) for financial support. We would like to thank the referee for his/her useful comments about the fits.
\end{acknowledgements}


\begin{thebibliography}{}
  
   \bibitem[Beltr\'an et al. (2009)]{2009ApJ...690L..93B} Beltr\'an, M.~T., 
Codella, C., Viti, S., Neri, R., \& Cesaroni, R.\ 2009, \apjl, 690, L93 

   \bibitem[Beltr\'an et 
al. (2005)]{2005A&A...435..901B} Beltr\'an, M.~T., Cesaroni, R., Neri, R., et al.\ 2005, \aap, 435, 901 

   \bibitem[Bennett 
\& Kaiser 2007]{2007ApJ...661..899B} Bennett, C.~J., \& Kaiser, R.~I.\ 2007, \apj, 661, 899

   \bibitem[Butler et al. (2001)]{2001ApJS..134..319B} Butler, R.~A.~H., De 
Lucia, F.~C., Petkie, D.~T., et al.\ 2001, \apjs, 134, 319  

   \bibitem[Carroll et al. (2010)]{2010ApJ...723..845C} Carroll, P.~B., Drouin, 
B.~J., \& Widicus Weaver, S.~L.\ 2010, \apj, 723, 845

   \bibitem[Cazaux et al. (2011)]{2011ApJ...741L..34C} Cazaux, S., Caselli, P., 
\& Spaans, M.\ 2011, \apjl, 741, L34 

   \bibitem[Ceccarelli et al. (1998)]{Ceccarelli1998} Ceccarelli, C.; Castets, .; Loinard, L.; Caux,E. and Tielens, A.G.G.M. 1998 A \& A, 338, L43

   \bibitem[Coutens et al. (2011)]{coutens2011} Coutens A., Vastel C., Caux E. et al. 2011, A \& A in press

   \bibitem[Hollis et al. (2000)]{2000ApJ...540L.107H} Hollis, J.~M., Lovas, 
F.~J., \& Jewell, P.~R.\ 2000, \apjl, 540, L107 

   \bibitem[Halfen et al. (2006)]{2006ApJ...639..237H} Halfen, D.~T., Apponi, 
A.~J., Woolf, N., Polt, R., \& Ziurys, L.~M.\ 2006, \apj, 639, 237 

   \bibitem[Hamberg et 
al. 2010]{2010A&A...514A..83H} Hamberg, M., {\"O}sterdahl, F., Thomas, R.~D., et al.\ 2010, \aap, 514, A83 

   \bibitem[Jalbout et al. 2007]{2007AsBio...7..433J} Jalbout, A.~F., Abrell, 
L., Adamowicz, L., et al.\ 2007, Astrobiology, 7, 433 

   \bibitem[Lattelais et al. 2009]{2009ApJ...696L.133L} Lattelais, M., 
Pauzat, F., Ellinger, Y., \& Ceccarelli, C.\ 2009, \apjl, 696, L133 

\bibitem[Margules et al. (2010)]{Margules2010} Margul\`es, L., Motiyenko, R.~A., Alekseev, E.~A.,  \& Demaison, J. 2010, J. Mol. Spectrosc., 260, 23

	\bibitem[Marstokk \& M{\o}llendal (1970)]{1970JMoSt...5..205M} Marstokk, K., \& M{\o}llendal, H.\ 1970, J. Mol. Struct., 5, 205 

   \bibitem[Marstokk \& M{\o}llendal (1971)]{1971JMoSt...7..101M} Marstokk, K. \& M{\o}llendal, H \ 1971, J. Mol. Struct., 7, 101

   \bibitem[Marstokk 
\& M{\o}llendal (1973)]{1973JMoSt..16..259M} Marstokk, K., \& M{\o}llendal, H.\ 1973, J. Mol. Struct., 16, 259

   \bibitem[Mauersberger et 
al. (1988)]{1988A&A...194L...1M} Mauersberger, R., Henkel, C., Jacq, T., \& Walmsley, C.~M.\ 1988, \aap, 194, L1 

   \bibitem[Motiyenko et al., 2010]{Motiyenko2010} Motiyenko, R.~A., Margul\`es, L., Alekseev, E.~A., Guillemin, J.~C., \& Demaison, J. 2010, J. Mol. Spectrosc., 264, 94
   
   \bibitem[Oliveira \& H\'ebrard, 2006]{Oliveira} Oliveira C. \& H\'ebrard G. 2006, \apj 653, 345
   
   \bibitem[Parise et al. (2002)]{Parise2002} Parise, B.; Ceccarelli, C.; Tielens, A.~G.~G.~M. et al. 2002 A \& A, 393, L49
   
   \bibitem[Parise et al. (2004)]{Parise2004} Parise, B.; Castets, A.; Herbst, E. et al. 2004, A \& A 416, 159
   
   \bibitem[Pickett et al. (1998)]{1998JQSRT..60..883P} Pickett, H.~M., 
Poynter, R.~L., Cohen, E.~A., et al.\ 1998, \jqsrt, 60, 883

   \bibitem[Ratajczak et 
al. (2011)]{2011A&A...528L..13R} Ratajczak, A., Taquet, V., Kahane, C., et al.\ 2011, \aap, 528, L13 

   \bibitem[Simakov et al. (2011)]{Simakov} Simakov, A., Sekiguchi, O., Bunkan, A., Joakim C., Uggerud, E.,\ 2011, J. Am. Chem. Soc, 133, 20816-20822

   \bibitem[Turner (1990)]{1990ApJ...362L..29T} Turner, B.~E.\ 1990, \apjl, 
362, L29 

   \bibitem[Widicus Weaver et al. (2005)]{2005ApJS..158..188W} Widicus Weaver, 
S.~L., Butler, R.~A.~H., Drouin, B.~J., et al.\ 2005, \apjs, 158, 188


\end{thebibliography}
\end{document}